\documentclass[conference, a4]{IEEEtran}
\usepackage{amsmath}
\usepackage{amssymb}
\usepackage{algorithmicx}
\usepackage{algpseudocode}
\usepackage{graphicx}
\usepackage{epstopdf}
\usepackage{cite}
\usepackage{textcomp}
\usepackage{mathrsfs}
\usepackage{color}
\usepackage{booktabs}
\usepackage{cases}
\usepackage{setspace}
\usepackage{bm}
\usepackage{cuted}
\usepackage{booktabs}
\usepackage{stfloats}
\usepackage{makecell}
\usepackage{mathtools}
\usepackage[ruled,linesnumbered]{algorithm2e}
\usepackage{graphicx}
\usepackage{caption}
\usepackage{subcaption} 

\makeatletter

\renewcommand{\maketag@@@}[1]{\hbox{\m@th\normalsize\normalfont#1}}%

\makeatother

\begin{document}
	
	\title{\huge Revisiting Wireless-Powered MEC: A Cooperative Energy Recycling Framework for Task-Energy Co-Design} 
    
\author{
	\IEEEauthorblockN{Haohao Qin\IEEEauthorrefmark{1}\IEEEauthorrefmark{2}, Bowen Gu\IEEEauthorrefmark{1}\IEEEauthorrefmark{2}, Xianhua Yu\IEEEauthorrefmark{3},  Hao Xie\IEEEauthorrefmark{3}, Yongjun Xu\IEEEauthorrefmark{4}, Qihao Li\IEEEauthorrefmark{5}, Liejun Wang\IEEEauthorrefmark{1}\IEEEauthorrefmark{2}}
	\IEEEauthorblockA{\IEEEauthorrefmark{1} School of Computer Science and Technology, Xinjiang University,  Urumqi, China}
    \IEEEauthorblockA{\IEEEauthorrefmark{2} Xinjiang Multimodal Intelligent Processing and Information Security Engineering Technology Research Center, Urumqi, China}
\IEEEauthorblockA{\IEEEauthorrefmark{3}  School of Electrical Engineering and Intelligentization, Dongguan University of Technology, Dongguan, China}
	\IEEEauthorblockA{\IEEEauthorrefmark{4} School of Communication and Information Engineering, Chongqing University of Posts and Telecommunications, Chongqing, China}
    \IEEEauthorblockA{\IEEEauthorrefmark{5} School of Communication Engineering, Jilin University, Jilin, China}
    \IEEEauthorblockA{\textit{Correspoding Author}: Bowen Gu (bwgu@xju.edu.cn)\vspace{-12pt}}
}

\setlength{\textfloatsep}{3pt}
\setlength{\abovedisplayskip}{2pt}
\setlength{\belowdisplayskip}{2pt}

	\maketitle
	\thispagestyle{empty}
	\pagestyle{empty}
	
	\begin{abstract}
    Cooperative energy recycling (CER) offers a new way to boost energy utilization in wireless-powered multi-access edge computing (MEC) networks, yet its integration with computation–communication co-design remains underexplored. This paper proposes a CER-enabled MEC framework that maximizes the minimum computable data among users under energy causality, latency, and power constraints. The intractable problem is reformulated into a convex form through relaxation, maximum ratio combining, and variable substitution, and closed-form solutions are derived via Lagrangian duality and alternating optimization, offering analytical insights. Simulation results verify that the proposed CER mechanism markedly increases total computable data while maintaining equitable performance across heterogeneous users.
	\end{abstract}
	\IEEEpeerreviewmaketitle
	\section{Introduction}
	The Internet of Things (IoT) has witnessed rapid development, driven by its capability to enable ubiquitous connectivity and intelligent interaction among people, devices, and environments \cite{iotbackground}. As the number of connected devices and the volume of data traffic grow exponentially, IoT applications increasingly demand real-time data processing with low latency and high reliability. To meet these requirements, multi-access edge computing (MEC) has emerged as a key enabler by offloading computation-intensive tasks from resource-constrained IoT devices to nearby edge servers, thereby significantly reducing communication latency and improving responsiveness \cite{9363323}.
    
    Despite these advances, the energy supply model remains a major bottleneck for sustainable IoT deployment. Conventional battery-powered devices suffer from limited lifetime and incur high maintenance costs, especially in large-scale or hard-to-reach deployments. These limitations severely hinder the scalability and long-term operability of MEC-enabled IoT networks, calling for more sustainable energy provisioning solutions. In recent years, wireless-powered communication networks (WPCNs) have emerged as a promising solution to the energy supply bottleneck in IoT systems. By enabling IoT devices to harvest energy from dedicated power sources (PSs) or ambient radio frequency (RF) signals, WPCNs support self-sustainable and battery-free device operation \cite{10589561}. This approach eliminates the high costs and logistical challenges associated with manual battery replacement, thereby significantly improving the scalability and operational lifespan of large-scale IoT deployments. Consequently, integrating WPCNs with MEC creates a new architectural paradigm that enhances the sustainability, scalability, and real-time responsiveness of next-generation IoT networks.

    To unlock the potential of wireless-powered MEC systems, recent studies have investigated the joint optimization of computation offloading and energy management under various scenarios and constraints. For example, \cite{8960510} studied a single-user setting and proposed an energy allocation strategy combined with partial task offloading. Building on this work, \cite{9698985} developed a hybrid offloading scheme supporting both binary and partial modes, which achieved superior performance under high signal-to-noise ratio (SNR) conditions. Leveraging emerging technologies, \cite{10301686} incorporated backscatter communication into wireless-powered MEC, enabling ultra-low-power task offloading and proposing an energy-efficient computation framework. To enhance wireless channel conditions and strengthen offloading security, \cite{10107791} integrated intelligent reflecting surfaces (IRS) into the system design. For greater deployment flexibility, \cite{10444003} employed unmanned aerial vehicles in a hybrid active–passive MEC, improving fairness in energy-efficient computation. Furthermore, \cite{10756636} considered a multi–access point configuration in which users dynamically selected offloading destinations based on their harvested energy levels.

   Although existing studies advance sustainable, battery-free IoT deployments, most still adopt conventional architectures in which devices harvest energy independently from reserved RF sources.  Such isolated designs often lead to severe energy imbalance in dense networks, where some nodes face shortages while others accumulate surplus energy, ultimately constraining scalability and overall efficiency.  In contrast, energy recycling (ER) offers a paradigm shift by enabling devices to harvest energy not only from dedicated sources but also from peer transmissions, reclaiming otherwise wasted energy and improving system-wide utilization. However, ER remains underexplored in WPCNs, with only a few initial efforts, such as backscatter-assisted ER \cite{9888066} and active ER with IRS support \cite{10083178}, demonstrating its potential. More importantly, existing ER research has largely focused on physical-layer energy transfer, paying little attention to the computation-centric challenges in MEC systems. Unlike conventional WPCNs that primarily aim to sustain wireless transmissions, wireless-powered MEC networks require the integrated management of energy and computation resources to balance performance and fairness. This raises a key question: how to design an ER-enabled MEC architecture that maximizes energy utilization while ensuring fair computation among heterogeneous users. To address this challenge, we propose a fairness-aware cooperative energy recycling (CER) framework that integrates inter-device energy sharing with unified scheduling of local computation and task offloading, thereby enhancing the synergy between communication and computation. In particular, the main contributions of this paper are summarized as follows.
	
	\begin{itemize}
		\item To evaluate the proposed protocol, we develop a resource management framework that maximizes the computable data by jointly optimizing communication and computation resources, while accounting for energy causality, latency, and power constraints, with a max–min fairness objective embedded in the optimization problem.
		\item To tackle this intractable problem, we first linearize the non-smooth objective via a relaxation method and adopt maximum ratio combining (MRC) to simplify the receive beamforming design. We then introduce variable substitution and employ alternating optimization to separate the interdependent variables. Leveraging Lagrangian duality theory, we derive closed-form solutions that provide analytical insights into system behavior and further quantify the performance gains brought by the proposed CER mechanism under specific configurations.
		\item Simulation results demonstrate that the proposed algorithm not only enhances the total computable data but also ensures equitable resource distribution among users.
	\end{itemize}

	\section{System Model}
    
	\subsection{System Structure and Transmission Mechanism} 
    
	A typical wireless-powered MEC system is considered, as shown in Fig. 1. The system comprises a set of $K$ wireless sensors (WSs), indexed by $k \in \mathcal{K} = \{1, 2, \cdots, K\}$, a multi-antenna access point (AP) equipped with $N$ receiving antennas, indexed by $n \in \mathcal{N} = \{1, 2, \cdots, N\}$, and a single-antenna PS. The AP is integrated with an MEC server to facilitate the remote execution of computational tasks. Each WS is assigned a computation task and is capable of harvesting energy to power both local execution and task offloading. To capture practical wireless conditions, we assume quasi-static flat-fading channels, where channel coefficients remain constant during a frame and vary independently across frames \cite{9887822}. The system operates over a time frame of duration $T$, divided into two phases: task offloading and edge computing. Due to the MEC server's strong processing capability and the small size of computation results, the time required for edge-side execution and result downloading is considered negligible, i.e., $\epsilon \approx 0$ \cite{10015648}.
    Local computing can be performed throughout the entire frame, while offloading is managed via time-division multiple access (TDMA), with each WS assigned a dedicated transmission time $t_k$ to avoid interference. Hence, the total transmission time satisfies, i.e., $\sum_{k=1}^K t_k \le T-\epsilon$.

       \begin{figure}[t]
		\centerline{\includegraphics[width=2.8in]{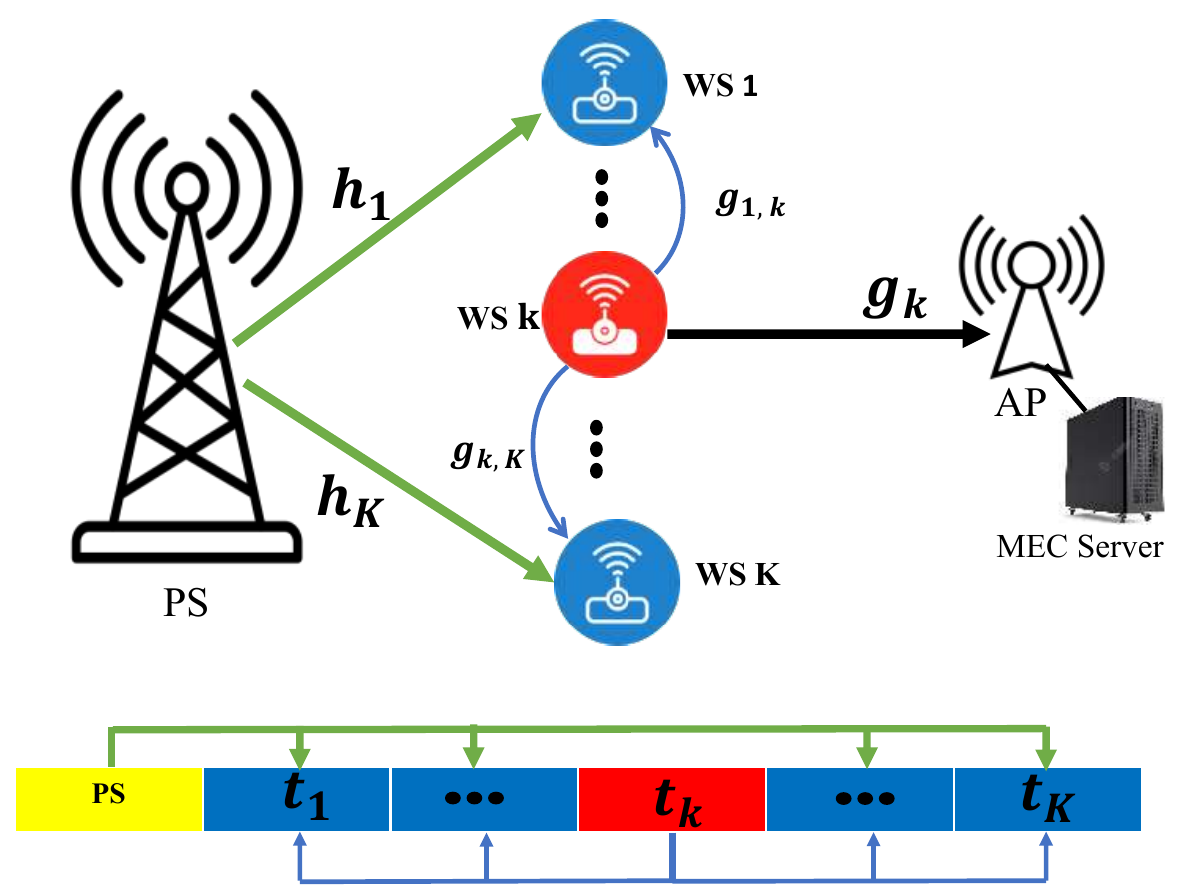}}
		\caption{A WPCN-assisted MEC system with energy recycling.}
		\label{fig1}
	\end{figure} 
    
	\subsection{ Energy Recycling Mechanism}
	
 Since single-antenna WSs cannot perform EH and data transmission at the same time, they harvest energy from the PS during their non-transmission slots \cite{10107791}.
    Moreover, they can recycle energy from the signals transmitted by other WSs during their respective transmission slots. The received signal at the $k$-th WS for EH can therefore be expressed as
	\begin{equation} \label{s1}
		y_{k}^{\text{Rx}}=\sum_{i=1,i\neq k}^K \sqrt{P_i} h_ks_i+\sum_{i=1,i\neq k}^K \sqrt{p_i} g_{i,k} x_i+ n_k,
	\end{equation}
where $h_k$ represents the channel coefficients between the PS and the $k$-th WS, and $g_{i,k}$ denotes the channel coefficients between the $i$-th WS and the $k$-th WS. $s_i$ and $x_i$ are the transmitted symbols from the PS and the $i$-th WS, respectively, satisfying $\mathbb{E}(|s_i|^2)=1$ and $\mathbb{E}(|x_i|^2)=1$. $n_k$ is the additive noise at the $k$-th WS. Additionally, $P_i$ and $p_i$ represent the transmit powers of the PS and the $i$-th WS, respectively, during the $i$-th slot. 
	
Since the contribution of the noise to the harvested energy is considered negligible \cite{10301686}, the total energy harvested by the $k$-th WS is expressed as
\begin{equation} \label{eh1} E_k^{\text{EH}}=\underbrace{\sum\limits_{i=1,i\neq k}^K \eta t_i P_i |h_k|^2}_{\text{Harvested from the PS}} + \underbrace{\sum\limits_{i=1,i\neq k}^K \eta t_i p_i |g_{i,k}|^2}_{\text{Recycled from other WSs}}, 
\end{equation} 
where $t_i$ denotes the transmission time allocated to the $i$-th WS, and $\eta \in (0, 1]$ denotes the energy conversion efficiency. 

\subsection{Task Execution Mechanism}

It is assumed that computational tasks are bitwise independent, supporting arbitrary partitioning of task data \cite{10486847}. Here, a partial offloading mechanism is considered, enabling the simultaneous execution of local computation and remote offloading.

\subsubsection{Local Computing} Let $C_k$ represent the number of central processing unit (CPU) cycles required to compute one bit of data. The number of bits computed locally by the $k$-th WS can be expressed as
\begin{equation} \label{lc1}
	R_k^{\text{LC}}=\frac{T f_k}{C_k},
\end{equation}
where $f_k$ denotes the CPU operating frequency of the $k$-th WS.

Each WS adopts an advanced dynamic voltage and frequency scaling (DVFS) technique \cite{9422161}. For analytical tractability, it is assumed that the CPU frequency $f_k$ remains fixed during each operational frame. Thus, the energy consumption for local computing at the $k$-th WS is given by
\begin{equation} \label{lc2}
	E_k^{\text{LC}}= T\phi_kf_k^3,
\end{equation}
where $\phi_k$ denotes the switched capacitance coefficient of the $k$-th WS.	

\subsubsection{Computation Offloading}  The signal received at the AP from the $k$-th WS during computation offloading is expressed as
\begin{equation} \label{s2}
	y_k^{\text{Tx}}=\bm w_k^{\text{H}} \bm g_k \sqrt{p_k} x_k+ \bm w_k^{\text{H}} \bm z_{\text{M}},
\end{equation}
where $\bm g_k \in \mathbb{C}^{N\times1}$ denotes the channel vector between the $k$-th WS and the AP, and $\bm z_{\text{M}}$ represents the noise at the AP, with $\bm z_{\text{M}} \sim \mathcal{CN}(0, \delta_{\text{M}}^2\bm I_N)$. Additionally, $\bm w_k \in \mathbb{C}^{N\times1}$ is the receive beamforming vector, satisfying $||\bm w_k||^2=1$.

Consequently, the data size offloaded by the $k$-th WS can be expressed as 
\begin{equation} \label{co1}
	R_k^{\text{CO}}=t_k B \log_2\left( 1+\dfrac{p_k|\bm w_k^{\text{H}}\bm g_k|^2}{\delta_{\text{M}}^2}\right) ,
\end{equation}
where $B$ denotes the system bandwidth allocated for computation offloading. 
	
Therefore, the total data size processed by the $k$-th WS, combining local computing and offloading, is given by
\begin{equation} \label{rt}
	R_k=R_k^{\text{LC}}+R_k^{\text{CO}}.
\end{equation}	

Moreover, the energy needed by the $k$-th WS for its task can be given by
\begin{equation} \label{et}
	E_k^{\text{EC}}=E_k^{\text{LC}}+E_k^{\text{CO}},
\end{equation}
where $E_k^{\text{CO}}=p_kt_k$ denotes the energy consumed by the $k$-th WS for computation offloading.

	\section{Problem Formulation and Algorithm Design}
	
	\subsection{Problem Formulation}
	
      In wireless-powered MEC networks, the coexistence of diverse computational demands and asymmetric energy availability across WSs naturally raises concerns about service disparity. Without careful coordination, weaker WSs may suffer from insufficient computing opportunities, leading to degraded user experience and unbalanced system performance. To address this, we adopt a max-min fairness criterion that aims to maximize the minimum amount of computable data across all WSs, ensuring fair access to MEC services. Accordingly, the optimization problem is formulated as follows
		\begin{equation}  \label{p1}  
	\begin{aligned}	
		& \underset{P_k, t_k, p_k, f_k, \bm w_k}{\mathop{\max }}\,  \underset{\forall k} {\mathop{\min }}\, \ R_k \\
		&\text{s.t.}~
		C_1: P_k \le P_{\max},  \forall k, ~C_2: \sum_{k=1}^K t_k \le T-\epsilon, \\
		&\quad ~~C_3: E_k^{\text{EC}}\le E_k^{\text{EH}}, \forall k, ~C_4: f_k \le f_k^{\max}, \forall k, \\
		&\quad ~~C_5: ||\bm w_k||^2=1, \forall k, ~C_6:  R_k \ge R_k^{\min}, \forall k,  \\
	\end{aligned}
\end{equation}  
 where  $P_{\max}$ denotes the maximum power supported by the PS, $f_k^{\max}$ is the maximum CPU frequency of the $k$-th WS, and $R_k^{\min}$ represents the minimum data size that the $k$-th WS must compute. In problem (\ref{p1}), $C_1$ restricts the PS transmit power within its hardware capability. $C_2$ ensures that the cumulative offloading durations across WSs do not exceed the available time budget, considering edge processing latency. $C_3$ imposes the energy causality condition, requiring that each WS’s total energy usage for computation and offloading remains within its harvested energy. $C_4$ bounds the local CPU frequency by each WS’s computational capacity. $C_5$ imposes unit-norm constraints on the receive beamforming vectors at the AP for proper signal processing. Finally, $C_6$ enforces a minimum data processing requirement of each WS to satisfy basic QoS expectations.
 
\subsection{Problem Transformation}

As can be seen, the inherent non-convexity of problem (\ref{p1}) arises from its non-smooth objective function and the strong coupling among transmission times $t_k$, transmit powers $p_k$ and $P_k$, and receive beamforming vectors $\bm w_k$. To facilitate tractability, we first introduce a slack variable $\gamma$ and impose the constraint $R_k \ge \gamma, \forall k$, which equivalently reformulates problem (\ref{p1}) into
		\begin{equation}  \label{p2}  
	\begin{aligned}	
		& \underset{P_k, t_k, p_k,  f_k, \gamma}{\mathop{\max }}\, \gamma  \\
		&\text{s.t.}~C_1\sim C_6, C_7: R_k \ge \gamma, \forall k. 
	\end{aligned}
\end{equation}

Then, we adopt MRC for receive beamforming, which maximizes the received SNR of each WS independently. Therefore, the receive beamforming vector of the $k$-th WS can be expressed as
\begin{equation} \label{m7}
	\bm w_n\triangleq \frac{\bm g_k}{||\bm g_k||}.\\
\end{equation}	

Substituting (\ref{m7}) into (\ref{rt}), the computable data size for the $k$-th WS is rewritten as
\begin{equation}
	\hat R_k =R_k^{\text{LC}}+t_kB\log_2\left( 1+\dfrac{p_k||\bm g_k||^2}{\delta_{\text{M}}^2}\right).
\end{equation}

After eliminating the dependency on $\bm{w}_k$, problem (\ref{p2}) remains challenging due to the coupling between transmission time $t_k$ and transmit powers $p_k$ and $P_k$. To address this, we introduce auxiliary variables $\bar{p}_k = p_k t_k$ and $\bar{P}_k = P_k t_k$, respectively. Hence, problem (\ref{p2}) can be reformulated into a more tractable form, i.e., 
		\begin{equation}  \label{p3}  
	\begin{aligned}	
		& \underset{\bar P_k, t_k, \bar p_k,  f_k, \gamma}{\mathop{\max }}\, \gamma  \\
		&\text{s.t.}~C_2, C_4,\bar C_1: \bar P_k \le P_{\max}t_k, \forall k,\\
		&\quad~~ \bar C_3: \bar E_k^{\text{EC}}\le \bar E_k^{\text{EH}}, \forall k, ~\bar C_6: \bar R_k \ge R_k^{\min}, \forall k,  \\
            &\quad ~~ \bar C_7:  \bar R_k \ge \gamma, \forall k, ~C_8: \bar P_k\ge 0, \bar p_k\ge 0, \forall k, \\          
	\end{aligned}
\end{equation}
where $\bar R_k =R_k^{\text{LC}}+t_kB\log_2\left( 1+\dfrac{\bar p_k||\bm g_k||^2}{t_k\delta_{\text{M}}^2}\right)$, $\bar E_k^{\text{EC}}=T\phi_kf_k^3+\bar p_k$, and $\bar E_k^{\text{EH}}=\sum\limits_{i=1,1\neq k}^K \eta \bar P_i |h_k|^2 + \sum\limits_{i=1,1\neq k}^K \eta \bar p_i |g_{i,k}|^2$.

\subsection{Algorithm Design}

\begin{table*} [t]
\small
\setcounter{equation}{13}
\begin{equation} \label{ape1}
	\begin{aligned}
		&\mathcal L_1 (\gamma, t_k, f_k, \lambda_1^k, \lambda_2, \lambda_3^k, \lambda_4^k,\lambda_5^k,\lambda_6^k)= \gamma 
		+\sum\limits_{k=1}^K\lambda_1^k(P_{\max}t_k{-}\bar P_k)
		+\lambda_2 (T-\epsilon-\sum_{k=1}^K t_k )
		+ \sum\limits_{k=1}^K\lambda_3^k(\bar E_k^{\text{EH}}
        -\bar E_k^{\text{EC}})\\
        &+\sum\limits_{k=1}^K\lambda_4^k(f_k^{\max}-f_k)
		+ \sum\limits_{k=1}^K\lambda_5^k (\bar R_k- R_k^{\min})+\sum\limits_{k=1}^K\lambda_6^k ( \bar R_k- \gamma),\\
	\end{aligned}
\end{equation}
\hrule 
\setcounter{equation}{16}
\begin{equation} \label{ape8}
	\begin{aligned}
		&\mathcal L_2 (\gamma, \bar P_k, \bar p_k, \varepsilon_1^k, \varepsilon_2^k, \varepsilon_3^k, \varepsilon_4^k, \varepsilon_5^k, \varepsilon_6^k )
		=\gamma
		+\sum\limits_{k=1}^K\varepsilon_1^k(P_{\max}t_k{-}\bar P_k)+ \sum\limits_{k=1}^K\varepsilon_2^k(\bar E_k^{\text{EH}}-\bar E_k^{\text{EC}})
		 + \sum\limits_{k=1}^K\varepsilon_3^k (\bar R_k- R_k^{\min})\\
         &+\sum\limits_{k=1}^K\varepsilon_4^k \bar P_k+\sum\limits_{k=1}^K\varepsilon_5^k \bar p_k
	+\sum\limits_{k=1}^K\varepsilon_6^k ( \bar R_k - \gamma),
	\end{aligned}
\end{equation}
\hrule
\end{table*}

\begin{algorithm}[t]
\small
	\caption{Max-min Fairness-Based Algorithm}
	\SetAlgoLined
	\KwIn{System parameters: $P_{\max}$, $T$, $\epsilon$, $f_k^{\max}$, $B$, $\delta_{\text{M}}^2$, etc.}
	\KwOut{Optimal solution $\{\bar P_k^*, t_k^*, \bar p_k^*, f_k^*, \gamma^*\}$.}
	
	Initialize $\{\bar P_k, t_k, \bar p_k, f_k, \gamma\}$ with feasible values;\\
	
	\Repeat{convergence of $\{\bar P_k, t_k, \bar p_k, f_k, \gamma\}$}{
		\textbf{Step 1:} Optimize $t_k$ and $f_k$ with fixed $\bar P_k$ and $\bar p_k$\\
		\quad Solve (\ref{ot2}) and (\ref{of2}) using the current Lagrange multipliers $\{\lambda_1^k, \lambda_2, \lambda_3^k, \lambda_4^k, \lambda_5^k, \lambda_6^k\}$.\\
		
		\textbf{Step 2:} Optimize $\bar p_k$ and $\bar P_k$ with fixed $t_k$ and $f_k$\\
		\quad Solve (\ref{op2}) and (\ref{obp}) using updated multipliers $\{\varepsilon_1^k, \varepsilon_2^k, \varepsilon_3^k, \varepsilon_4^k, \varepsilon_5^k, \varepsilon_6^k\}$.\\
		
		\textbf{Step 3:} Update slack variable $\gamma$\\
		\quad Set $\gamma = \min_{\forall k} \bar R_k(\bar P_k, t_k, \bar p_k, f_k)$.
	}
\end{algorithm}

Although problem (\ref{p3}) is convex, obtaining closed-form optimal solutions remains challenging due to the residual coupling among decision variables. To resolve this, we employ an alternating optimization framework: in each iteration, the transmission time and CPU frequency are jointly optimized while fixing the transmit powers at the PS and WSs, followed by an update of the transmit powers with the transmission time and CPU frequency held constant. This decoupling enables each subproblem to be solved in closed form. The overall procedure is outlined in the proposed max–min fairness-based algorithm (MFBA) in \textbf{Algorithm~1}, with the closed-form solutions for both subproblems derived as follows.

Specifically, for fixed transmit powers ${\bar P_k}$ and ${\bar p_k}$, the Lagrangian function associated with the optimization of $t_k$ and $f_k$ is given in (\ref{ape1}), where $\lambda_1^k$, $\lambda_2$, $\lambda_3^k$, $\lambda_4^k$, $\lambda_5^k$, and $\lambda_6^k$ denote the non-negative Lagrange multipliers corresponding to the respective constraints. By taking the first-order derivatives with respect to $t_k$ and $f_k$ and applying the Karush–Kuhn–Tucker (KKT) conditions \cite{Boyd_Vandenberghe_2004}, the optimal solutions are obtained in closed form as
\setcounter{equation}{14}
\begin{equation} \label{ot2}
	t_k^*= \dfrac{\bar p_k||\bm g_k||^2}{\delta_{\text{M}}^2\mathcal{F}^{-1}\left( \dfrac{\lambda_2-\lambda_1^kP_{\max}}{(\lambda_5^k+\lambda_6^k) B}\right)}, \forall k,
\end{equation}
\begin{equation} \label{of2}
	f_k^*= \sqrt{\dfrac{(\lambda_5^k+\lambda_6^k)\frac{ T}{C_k}-\lambda_4^k}{3\lambda_3^k T\phi_k}},  \forall k,
\end{equation}
where $\mathcal{F}(x) = \frac{1}{\ln 2} \left[ \ln(1+x) - \frac{x}{1+x} \right]$ and
$\mathcal{F}^{-1}(x) = e^{W\left(-\frac{1}{e^{1+x\ln 2}}\right) + 1 + x\ln 2} - 1$,
with $W(\cdot)$ denoting the Lambert-$W$ function.

Next, given $t_k$ and $f_k$, the Lagrangian function for optimizing $\bar P_k$ and $\bar p_k$ is formulated in (\ref{ape8}), where $\varepsilon_1^k$, $\varepsilon_2^k$, $\varepsilon_3^k$, $\varepsilon_4^k$, $\varepsilon_5^k$, and $\varepsilon_6^k$ denote the non-negative Lagrange multipliers associated with the respective constraints. By differentiating the Lagrangian with respect to $\bar P_k$ and $\bar p_k$ and applying the KKT optimality conditions, the closed-form optimal solutions are obtained as
\setcounter{equation}{17}
\begin{equation} \label{op2}
	\bar p_k^*= \left[ \dfrac{(\varepsilon_3^k+\varepsilon_6^k)Bt_k||\bm g_k||^2}{\ln 2(\varepsilon_2^k-\varepsilon_5^k)}-\frac {t_k\delta_{\text{M}}^2}{||\bm g_k||^2}\right]^+,  \forall k,
\end{equation}
\begin{equation} \label{obp}
		\bar P_k^*= [P_{\max}t_k]^+,  \forall k,
\end{equation}
where $[x]^+=\max\{0,x\}$.

\textbf{\textit{Remark 1}}: It can be observed from (\ref{ot2}) that $t_k^*$ follows a time-domain water-filling structure, whereby users with stronger channels $||\bm g_k||^2$ naturally obtain longer offloading durations. This advantage, however, is deliberately moderated by the fairness-related multiplier $\lambda_6^k$, which reallocates part of the time budget from strong users to weaker ones.
From (\ref{of2}), $f_k^*$ grows proportionally to the square root of available resources under the cubic CPU power model, enabling higher local execution frequencies with increased resources, while $\lambda_6^k$ similarly restrains excessive allocation to ensure weaker WSs are not deprived.  (\ref{op2}) adopts a threshold-based activation rule, whereby a WS engages in offloading only if its channel gain and marginal utility exceed the composite “energy–fairness price” reflected in the dual variables; otherwise, it defaults to local execution. Meanwhile, (\ref{obp}) indicates that the PS should always transmit at $P_{\max}$ during each allocated slot, an outcome of the linear EH model and the absence of inter-slot average power coupling, thus maximizing instantaneous energy transfer.
Overall, these results demonstrate that while favorable channels still lead to greater resource allocations, the fairness multipliers actively suppress disproportionate resource domination by strong WSs. This makes the proposed fairness-aware CER network particularly suitable for scenarios where edge users or weak-signal devices must be guaranteed a minimum operational capability, even under resource scarcity.

\subsection{Offloading capacity enhancement with CER}

To analytically isolate and quantify the impact of CER on offloading capacity, we consider a simplified scenario in which each WS adopts a full-offloading strategy, i.e., all harvested energy is devoted solely to data transmission. To eliminate scheduling effects and focus on inter-WS energy recycling, an equal time division policy and a constant PS transmit power are assumed, namely $t_k = 1/K$ and $P_k = P_0$ for all $k$. Under these assumptions, the offloaded data sizes with and without CER can be expressed as
\begin{equation} \label{sc1}
	R_k^{\text{w. ER}}= \frac{1}{K} \log_2 \left(1+ \dfrac{K||\bm g_k||^2 E_k^{\text{w. ER}}}{\delta_{\text{M}}^2} \right), 
\end{equation} 
\begin{equation} \label{sc2}
	R_k^{\text{w.o. ER}}= \frac{1}{K} \log_2 \left(1+ \dfrac{K||\bm g_k||^2 E_k^{\text{w.o. ER}}}{\delta_{\text{M}}^2} \right), 
\end{equation} 
where 
\begin{equation}
	E_k^{\text{w. ER}}= \sum\limits_{i=1,1\neq k}^K \frac{1}{K} \eta   P_0 |h_k|^2+\sum\limits_{i=1,1\neq k}^K \frac{1}{K} \eta  p_i |g_{i,k}|^2,
\end{equation}
and 
\begin{equation}
	E_k^{\text{w.o. ER}}=\sum\limits_{i=1,1\neq k}^K  \frac{1}{K}  \eta P_0 |h_k|^2.
\end{equation}

The gap in offloading capacity between the two schemes can be expressed as (\ref{sc3}), 
where approximation ($a$) assumes $\delta_{\text{M}}^2 \approx 0$, given that the noise power is negligible compared to the transmit signal. 

\begin{table*}[t]
\small
\setcounter{equation}{23}
\begin{equation}  \label{sc3}
	\begin{aligned}
		R_k^{\text{Gap}}& =R_k^{\text{w. ER}}-R_k^{\text{w.o. ER}}
		=\frac{1}{K} \log_2\left(\frac{\delta_{\text{M}}^2+K||\bm g_k||^2E_k^{\text{w. ER}}}{\delta_{\text{M}}^2+K||\bm g_k||^2E_k^{\text{w.o. ER}}}\right)
	\overset{(a)}{\approx} \frac{1}{K}  \log_2\left( \frac{E_k^{\text{w. ER}}}{E_k^{\text{w.o. ER}}} \right)
	= \frac{1}{K} \log_2\left( 1+ \frac{ \sum\limits_{i=1,1\neq k}^K  p_i |g_{i,k}|^2}{\sum\limits_{i=1,1\neq k}^K  P_0 |h_k|^2} \right).\\		
	\end{aligned} 
\end{equation}
\hrule 
\end{table*}

\begin{figure*}[t]
\centering
\subfloat[Total data vs $P_{\max}$]{
    \includegraphics[width=0.24\textwidth]{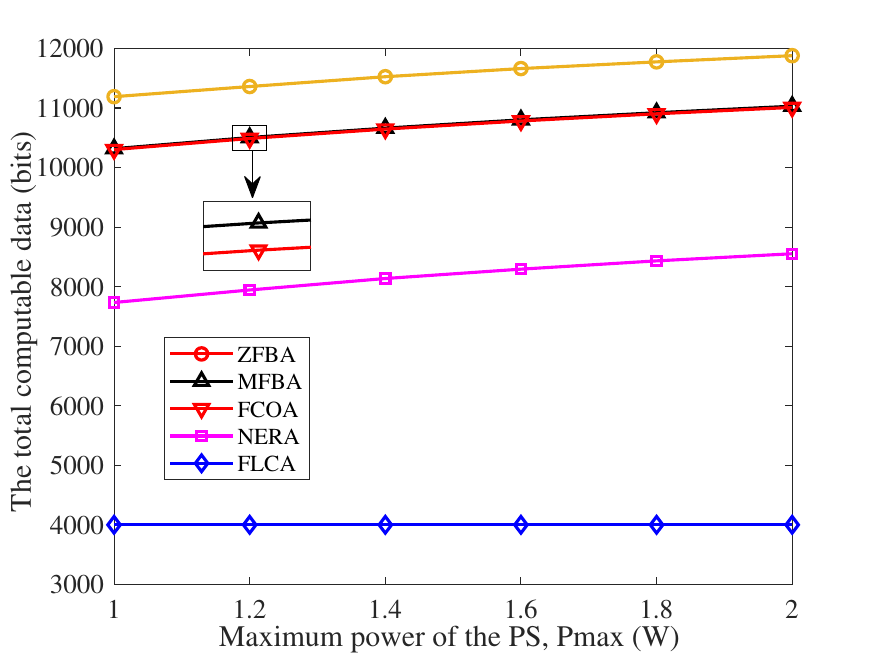}
}
\subfloat[Total data vs $K$]{
    \includegraphics[width=0.24\textwidth]{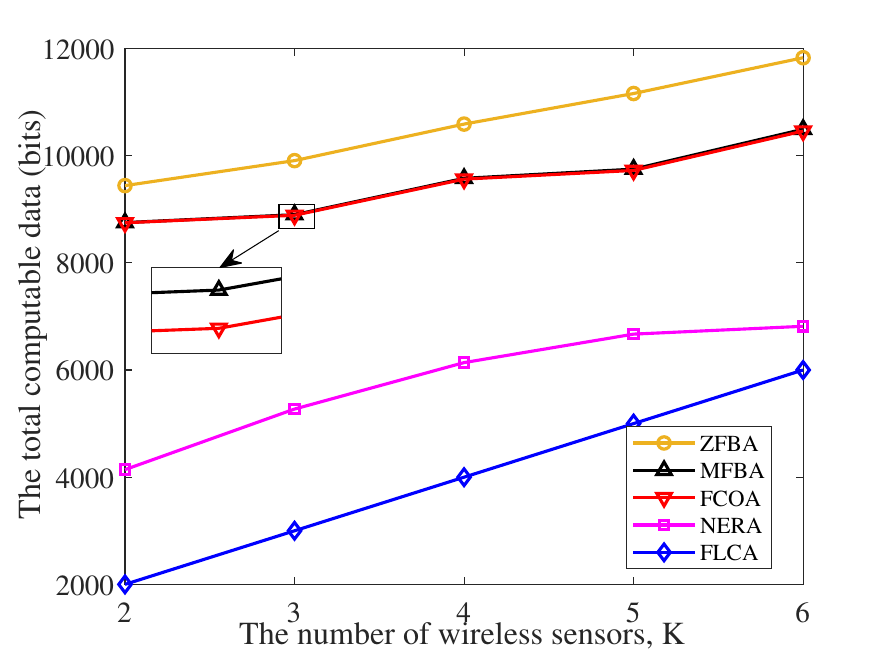}
}
\subfloat[Total data vs $N$]{
    \includegraphics[width=0.24\textwidth]{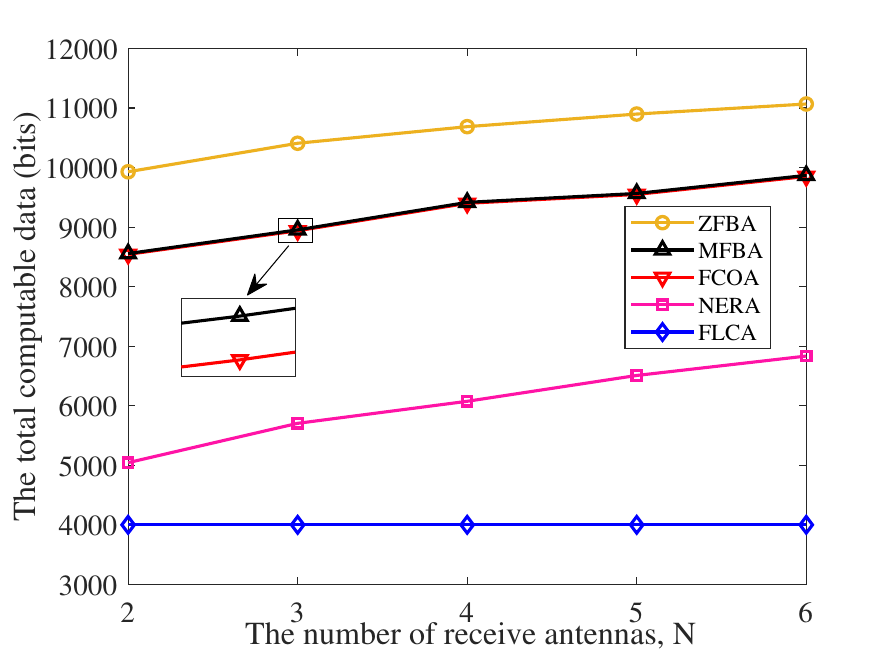}
}
\subfloat[Max/min data vs $K$]{
    \includegraphics[width=0.24\textwidth]{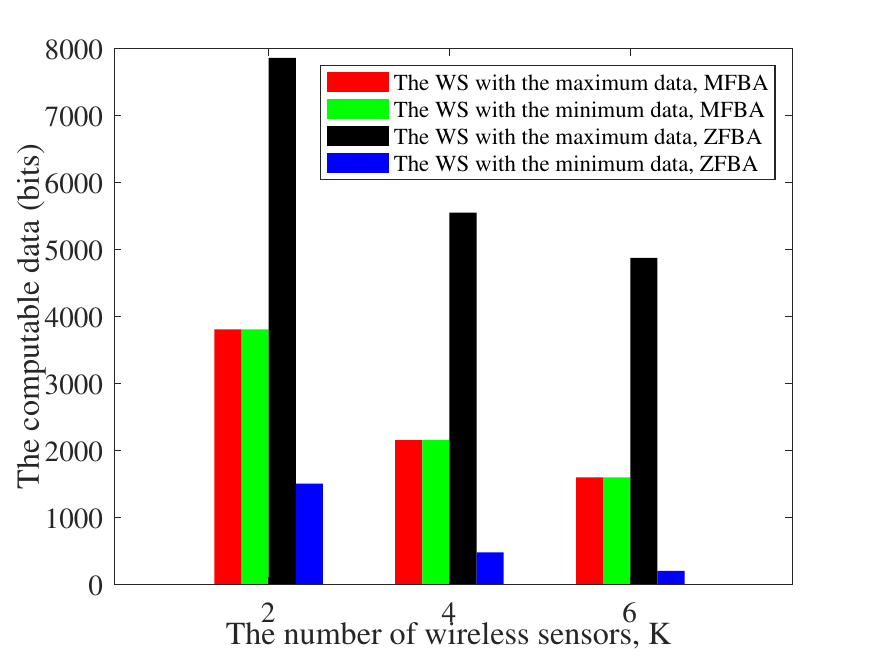}
}
\caption{Performance evaluation of the proposed algorithm: (a) total computable data under different $P_{\max}$; (b) total computable data under different number of WSs $K$; (c) total computable data under different number of receive antennas $N$; (d) maximum and minimum computable data versus $K$.}
\label{fig2}
\end{figure*}

 \textbf{\textit{Remark 2}}: From (\ref{sc3}), it can be observed that the performance gain $R_k^{\text{gap}}$ is mainly influenced by the inter-WS channel gain $g_{i,k}$ and the transmit power $p_i$ of WSs. When $g_{i,k}$ is small, the recyclable energy is limited, leading to a reduced gap between the CER-enabled and non-CER schemes. Conversely, stronger inter-WS links or higher $p_i$ values significantly enlarge this gap by creating more energy harvesting opportunities among WSs. Moreover, the improvement brought by CER is more evident when the direct PS–WS link $P_0|h_k|^2$ is weak, highlighting its role in enhancing offloading throughput under energy-limited conditions. These results suggest that CER can be particularly beneficial in dense WS deployments or scenarios with constrained PS transmission power, which will be further confirmed in the simulation section.

\section{Simulation Results}

In this section, we evaluate the proposed algorithm through numerical simulations. The simulated WPCN-assisted MEC system consists of one PS, one AP equipped with 4 receive antennas, and four single-antenna WSs. The distance between the PS and each WS, as well as between each WS and the AP, is within 15 m, while the inter-WS distance is within 5 m. Large-scale fading is modeled by distance-dependent path loss with an exponent of 2.2, and all channels experience Rayleigh small-scale fading \cite{10486847}. Unless otherwise stated, the system parameters follow \cite{10301686}:  $\eta=0.8$, $T=1$ s, $C_k=1000$ cycles/bit, $\phi_k=10^{-30}$, $\delta_{\text{M}}^2=-90$ dBm, $B=1$ KHz, $P_{\max}=1$ W, $f_k^{\max}=1$ MHz, $R_k^{\min}=100$bits.

Moreover, to evaluate the performance of the proposed MFBA algorithm, four benchmark schemes are considered for comparison: (i) the zero-fairness-based algorithm (ZFBA), (ii) the full local computing algorithm (FLCA), (iii) the full computation offloading algorithm (FCOA), and (iv) the non-cooperative ER algorithm (NERA).

Figs. \ref{fig2}~(a)-(c) jointly illustrate the influence of the power budget $P_{\max}$, the number of WSs $K$, and the number of AP antennas $N$ on the total computable data. Across all settings, FLCA exhibits slow, almost flat growth due to its reliance on purely local execution, which is fundamentally constrained by CPU frequency and latency limits. In contrast, all offloading-enabled schemes achieve notable gains. Among them, ZFBA consistently attains the highest throughput by allocating resources to the strongest WSs, whereas MFBA trades part of this advantage for fairness guarantees, leading to slightly reduced performance. Nevertheless, by jointly optimizing local computation and offloading while exploiting CER to redistribute surplus energy among WSs, MFBA delivers substantial gains over FCOA and NERA.

Looking more closely, Fig. \ref{fig2}~(a) shows that increasing $P_{\max}$ boosts harvested energy, thereby improving both local processing and offloading performance. Besides, CER-assisted algorithms gain disproportionately from this improvement thanks to peer energy transmission. Fig. \ref{fig2}~(b) reveals that increasing $K$ enhances multi-user diversity and scheduling flexibility, with the benefits of CER becoming increasingly pronounced in dense deployments. However, the fairness of MFBA limits its growth relative to ZFBA, which can fully exploit strong WSs. In Fig. \ref{fig2}~(c), increasing $N$ strengthens signal reception and improves offloading quality, producing steady throughput gains for all offloading-enabled algorithms without widening fairness disparities, as indicated by the parallel trends of MFBA and ZFBA. Overall, these results confirm the consistent contribution of CER to throughput improvement, with its relative advantages most evident in dense networks or under tight resource constraints.

As shown in Fig. \ref{fig2}~(d), increasing $K$ leads to a decline in both the maximum and minimum computable data of each WS for MFBA and ZFBA. This decrease stems from tighter resource partitioning and increased competition among WSs, which reduces the average resource share available to each node and thus limits individual computational throughput. Despite this general trend, MFBA maintains a consistently small gap between the best- and worst-performing WSs, indicating that its fairness-oriented allocation strategy remains robust and scalable across varying network sizes. In contrast, the differences exhibited by ZFBA when $K$ increases remain significant, reflecting its inherent bias toward high-efficiency WSs to maximize total system throughput. These observations underscore a key tradeoff: while ZFBA delivers higher aggregate performance, it does so at the cost of significant fairness loss, resulting in highly uneven user experiences. MFBA, on the other hand, achieves strong fairness guarantees with more graceful performance scaling, making it particularly suitable for dense deployments or scenarios where equitable service delivery is critical.

\section{Conclusions}

This paper investigated a fairness-aware CER architecture for wireless-powered MEC networks, jointly optimizing local computing and task offloading to maximize the minimum computable data among WSs. By linearizing the objective, simplifying beamforming via MRC, and applying variable substitution with alternating optimization, closed-form solutions were derived using Lagrangian duality, along with an analytical evaluation of CER gains. Simulations confirm that the proposed scheme significantly improves both total computable data and fairness compared with conventional benchmarks.

\section*{Acknowledgments}
This work was supported in part by the 2025 Xinjiang Tianchi Talents Young Doctors Fund under Grant No. 51052501823, in part by the Young Scientists Fund of the Natural Science Foundation of Xinjiang under Grant No. 2025D01C294, in part by the Central Guidance Local Science and Technology Development Fund under Grant No. ZYYD2025JD10, in part by the Guangdong Basic and Applied Basic Research Foundation under Grant 2024A1515110036, and in part by the National Natural Science Foundation of China under Grant 62472368, U23A20279, and 62271094.

\end{document}